\documentclass[journal]{IEEEtran}
\usepackage{stmaryrd}
\usepackage{amsfonts}
\usepackage{graphicx,times,amsmath}
\usepackage{graphics,color}
\usepackage{graphicx}
\usepackage{latexsym}
\usepackage{amsmath}
\usepackage{amsfonts}
\usepackage{amssymb}
\usepackage{url}
\usepackage{subfigure}
\usepackage{times}
\usepackage{color}
\usepackage{float}
\newtheorem{prop}{Proposition}
\newtheorem{defn}{\indent Deinition}

\newtheorem{exa}{\indent Example}
\newtheorem{re}{\indent Remark}

\allowdisplaybreaks

\title{On Synchronization: Comments on the paper "Synchronization in scale-free dynamical networks: robustness and
fragility", IEEE Trans. Circuits Syst. I 49 (1) (2002) 54-62}

\author{Tianping~Chen
\thanks{Tianping Chen is with the School of Computer/Mathematical
Sciences, Fudan University, 200433, Shanghai, P.R. China. Corresponding
author: Tianping Chen. E-mail:
tchen@fudan.edu.cn}
\thanks{This work was supported by
the National Science Foundation of China under Grant No. 61203149.}
}
\begin{document}
\maketitle

\begin{abstract}
Synchronization problem for linear coupled networks is a hot topic in
recent decade. However, until now, some confused concepts and results still
puzzle people. To avoid further misleading researchers, it is necessary to
point out these misunderstandings, correct these mistakes and give precise
results.

\end{abstract}

\section{Introduction}
In discussing synchronization of coupled systems, following concepts are
most important and should be addressed precisely:

\begin{enumerate}
\item
What is the synchronization and what is the synchronized state?

\item
Can an individual trajectory $\dot{s}(t)=f(s(t))$ of the uncoupled system
be the synchronized state of the coupled system?

\item
What is the relationship between the stability of a trajectory of the
uncoupled system and the stability of the synchronized state of the coupled
system;

\item

synchronization criteria of the coupled system.
\end{enumerate}

In \cite{Wang}, the authors wrongly consider the synchronization of the
coupled system as the stability of an individual trajectory of the
uncoupled system. Based on this misunderstanding, the authors define the so
called synchronized state inappropriately. Two criteria for the exponential
stability of the so called synchronized state are given. Unfortunately,
these two criteria are incorrect, too.

In this paper, we address this issue in detail, pointing out why the
results given in \cite{Wang} are incorrect. Furthermore, we clarify the
differences and relations among the stability of the trajectory of
uncoupled system, stability of the trajectory of coupled system and the
synchronization of coupled system.




\section{Comments on \cite{Wang}}
In the paper \cite{Wang}, the authors discussed the following coupled
networks
\begin{eqnarray}
\dot{x}^{i}(t)=f(x^{i}(t))+c\sum\limits_{j=1}^{N}a_{ij}\Gamma
x^{j}(t)\quad i=1,\cdots,N \label{synn2w}
\end{eqnarray}
and its synchronization. Here, $A=[a_{ij}]_{i, j=1}^{m}\in R^{N\times
N},~a_{ij}\ge 0,~i\ne j,~a_{ii}=-\sum_{j\ne i}a_{ij}$ and assumed to be
strongly connected, $\Gamma=diag[\gamma_{1},\cdots,\gamma_{n}]$.

The authors wrote in \cite{Wang}:

{\it Hereafter, the dynamical network is said to achieve (asymptotical)
synchronization if as
\begin{align}
x^{1}(t)=x^{2}(t)=\cdots=x^{N}(t)=s(t),~~t\rightarrow\infty
\label{wangs}
\end{align}
where $s(t)\in R^{n}$ is a solution of an isolate node, namely
\begin{align}
\dot{s}(t)=f(s(t))
\end{align}

Here, $s(t)$ can be an equilibrium point, a periodic orbit, or a chaotic
attractor. Clearly, stability of the synchronized states (\ref{wangs}) of
network (\ref{synn2w}) is determined by the dynamics of an isolate node
(function $f$ and solution $s(t)$), the coupling strength $c$, the inner
linking matrix $\Gamma$, and the coupling matrix $A$.}

First of all, mathematically, expression (\ref{wangs}) is meaningless. 

It is our understanding that the authors want to say
\begin{align}
\lim_{t\rightarrow\infty}(x^{i}(t)-s(t))=0,~~i=1,\cdots,N
\end{align}

Following lemmas (main results)  are given in \cite{Wang}, too.

{\it Lemma 1. Consider the dynamical network (\ref{synn2w}). Let
\begin{align}
0=\lambda_{1}>\lambda_{2}\ge \lambda_{3}\ge\cdots\ge \lambda_{N}
\end{align}
be the eigenvalues of its coupling matrix $A$. If the following of
$(N-1)$-dimensional linear time-varying systems
\begin{align}
\dot{w}(t)=(Df(s(t))+c\lambda_{k}\Gamma)w(t)~~~k=2, \cdots, N
\label{csaw}
\end{align}
are exponentially stable, then the synchronized states (\ref{wangs})
are exponentially stable.

If $s(t)=\bar{s}$ is an equilibrium point, then a necessary and sufficient
condition for the synchronization stability is that the real parts of the
eigenvalues of the matrix $[Df(\bar{s})+c\lambda_{2}\Gamma]$ are all
negative.}

{\it Lemma 2. Consider the network (\ref{synn2w}). Suppose that there exists
an $n\times n$ diagonal matrix $D>0$ and two constants $\tau>0$
and $\bar{d}<0$, such that
\begin{align}
[Df(s(t)+d\Gamma]^{T}D+D[Df(s(t)+d\Gamma]\le -\tau I_{n}
\end{align}
for all $d<\bar{d}$. If
\begin{align}
c\lambda_{2}\le \bar{d}
\end{align}
then the synchronized states (6) are exponentially stable.}

Unfortunately, the claims given in two lemmas are incorrect. 

In the following, we give detail explanations.

Denote $x(t)=[{x^{1}}^{\top}(t), \cdots, {x^{N}}^{\top}(t)]^{\top}\in
R^{nN}$, $S(t)=\newline[{s}^{\top}(t), \cdots, {s}^{\top}(t)]^{\top}\in R^{nN}$,
where $s(t)$ is a solution satisfying $\dot{s}(t)=f(s(t))$, and
$F(x(t))=[f(x^{1}(t))^{T}, \cdots, f(x^{N}(t))^{T}]^{\top}$,~ then the
system (\ref{synn2w}) can be written as
\begin{eqnarray}
\dot{x}=F(x(t))+c\left(A\otimes\Gamma\right)x(t)\label{chen}
\end{eqnarray}
and the asymptotical (exponential) stability of the synchronized state
$s(t)$ is equivalent to that $S(t)$ is an asymptotically
(exponentially) stable solution of (\ref{chen}).

Let $\delta x(t)$ be the variation near $S(t)$, then
\begin{eqnarray}
\dot{\delta}x(t)=[I_{N}\otimes DF(s(t))]\delta
x(t)+c\left(A\otimes\Gamma\right)\delta x(t)\label{chen1}
\end{eqnarray}
Moreover, write the Jordan decomposition as $A=\Phi^{\top} \Lambda
\Phi$,~$\Lambda=diag[\lambda_{1},\cdots,\lambda_{N}]$, where
$0=\lambda_{1}>\lambda_{2}\ge\cdots\ge \lambda_{N}$, and $\delta
u(t)=\Phi\delta x(t)=[\delta u^{1}(t)^{\top},\cdots,\delta
u^{N}(t)^{\top}]^{\top}$, then
\begin{eqnarray}
\dot{\delta} u(t)=[I_{N}\otimes DF(s(t))]\delta
u(t)+c\left(\Lambda\otimes\Gamma\right)\delta u(t)\label{chen1a}
\end{eqnarray}
which also can be written as
\begin{eqnarray}
\dot{\delta} u^{k}(t)=[D f(s(t))+\lambda_{k}\Gamma]\delta
u^{k}(t),~~k=1,\cdots,N \label{chen1aa}
\end{eqnarray}

Thus, the asymptotical (exponential) stability of the trajectory $s(t)$
with respect to the coupled system (\ref{synn2w})  is equivalent to the all
following "N" (\underline{not $N-1$}) equations
\begin{align}
\dot{w}(t)=(Df(s(t))+c\lambda_{k}\Gamma)w(t)~~~k=1, \cdots, N
\end{align}
are asymptotically (exponentially) stable.

Therefore, Lemma 1 and Lemma 2 in \cite{Wang} should be

{\it Lemma 1*. Consider the dynamical network (\ref{synn2w}). Let
\begin{align}
0=\lambda_{1}>\lambda_{2}\ge \lambda_{3}\ge\cdots\ge \lambda_{N}
\end{align}
be the eigenvalues of its coupling matrix $A$. If the following of
$(N)$-dimensional linear time-varying systems
\begin{align}
\dot{w}(t)=(Df(s(t))+c\lambda_{k}\Gamma)w(t)~~~k=1, \cdots, N
\end{align}
are exponentially stable, then the synchronized states (\ref{wangs}) are
exponentially stable.

{\it Lemma 2*. Consider the network (\ref{synn2w}). Suppose that there
exist an $n\times n$ diagonal matrix $D>0$ and a constant $\tau>0$, such
that
\begin{align}
[Df(s(t)]^{T}D+D[Df(s(t)]\le -\tau I_{n}
\end{align}
then the synchronized states (6) are exponentially stable.}

Furthermore, we can prove

{\it Lemma 1**. In case $\Gamma=I_{n}$, then the synchronized states
(\ref{wangs}) are asymptotically (exponentially) stable for the coupled
system (\ref{synn2w}), it is necessary and sufficient that the uncoupled
system
\begin{eqnarray}
\dot{w}(t)=D(f(s(t)))w(t)\label{Lemma 1**}
\end{eqnarray}
is asymptotically (exponentially) stable itself.

If $s(t)=\bar{s}$ is an equilibrium point, then a necessary and sufficient
condition for the synchronization stability is that the real parts of the
eigenvalues of the matrix $Df(\bar{s})$ are all negative.}

In fact, any solution of
\begin{align*}
\dot{\delta}x(t)=[I_{N}\otimes DF(s(t))+A\otimes I_{n}]\delta x(t)
\end{align*}
can be written as $\delta x(t)=e^{[A\otimes I_{n}]t}\delta x^{*}(t)$. Here,
$\delta x^{*}(t)$ satisfies the variational system near $S(t)$
\begin{eqnarray}
\dot{\delta}x^{*}(t)=[I_{N}\otimes DF(s(t))]\delta x^{*}(t)
\label{chen2}
\end{eqnarray}
and is asymptotically (exponentially) stable

From the asymptotical stability of (\ref{Lemma 1**}), we have
\begin{eqnarray*}
\lim_{t\rightarrow \infty}\delta x^{*}(t)=0
\end{eqnarray*}
combining with $\delta
x(t)=e^{[A\otimes I_{n}]t}\delta x^{*}(t)$ gives
\begin{eqnarray*}
\lim_{t\rightarrow \infty}\delta x(t)=0
\end{eqnarray*}
which implies
\begin{eqnarray*}
\lim_{t\rightarrow \infty}(x(t)-S(t))=0
\end{eqnarray*}
and equivalently,
\begin{eqnarray*}
\lim_{t\rightarrow \infty}(x^{i}(t)-s(t))=0,~~i=1,\cdots,N
\end{eqnarray*}

\begin{re}
It should be noted that in Lemma 1**, the condition $\Gamma=I_{n}$ plays
key role in the proof. In case that
$\Gamma=diag[\gamma_{1},\cdots,\gamma_{n}]$ with some $\gamma_{i}\ne
\gamma_{j}$, it is not yet known whether Lemma 1** is still true. The point
is $Df({s}(t))\Gamma\ne \Gamma Df({s}(t))$.

Similarly, in case $s(t)=\bar{s}$ is an equilibrium point, even the real
parts of the eigenvalues of the matrix $[Df(\bar{s})+c\lambda_{2}\Gamma]$
are all negative, we still can not derive the real parts of the eigenvalues
of the matrix $[Df(\bar{s})+c\lambda_{k}\Gamma]$, $k=3,\cdots,N,$ are all negative,
which also means that it is not yet known whether 
linear time-varying systems
\begin{align}
\dot{w}(t)=(Df(\bar{s})+c\lambda_{k}\Gamma)w(t)~~~k=3, \cdots, N
\end{align}
are exponentially stable. Therefore, the claim made in Lemma 1 of
\cite{Wang}: {\it if $s(t)=\bar{s}$ is an equilibrium point, then a
necessary and sufficient condition for $\bar{s}$ being stable is  that the
real parts of the eigenvalues of the matrix
$[Df(\bar{s})+c\lambda_{2}\Gamma]$ are all negative
 } is incorrect.
\end{re}

In the following, we will give a precise description of synchronization and
correct results.

\begin{defn} 
Synchronization subspace is the set composed of ${\mathcal
S}=\{(x^{{1}^{\top}}, \cdots, x^{{m}^{\top}})^{\top}:~x^{i}=x^{j}, i, j=1,
\cdots, m\}$, where $x^{i}=[x^{i}_{1}, \cdots, x^{i}_{n}]^{\top}\in
R^{n}$,~~$i=1, \cdots, m$.
\end{defn}

\begin{defn} 
(Local synchronization see \cite{Wu,Lu1,Lu2}) If for some $\delta>0$, such
that in case the distance between $x(t)$ and ${\mathcal S}$ at time $0$,
$d(x(0),\mathcal S)\le \delta $, we have
\begin{eqnarray*}
\lim_{t\rightarrow\infty}d(x(t),\mathcal S)=0,  ~~i, j=1, 2, \ldots, m
\end{eqnarray*}
Then, Synchronization subspace/manifold is local stable with respect to the
coupled system (\ref{synn2w}), or system (\ref{synn2w}) realizes local
synchronization.
\end{defn}

Denote (for asymmetric coupling matrix case, see \cite{Lu2})
$\bar{x}(t)=\frac{1}{N}\sum_{i=1}^{N}x^{i}(t),$
$\bar{X}(t)=[\bar{x}^{T}(t),\cdots,\bar{x}^{T}(t)]^{T}\in \mathcal{S}$.
$S(t)=[s^{T}(t),\cdots,s^{T}(t)]^{T}\in \mathcal{S}$, where $s(t)$
satisfies $\dot{s}(t)=f(s(t))$. $\delta \bar{x}(t)=x(t)-\bar{X}(t)$ is the
component in the transverse subspace.

From Figure 1, it can be seen that
synchronization means that  the component in the transverse subspace
$\delta \bar{x}(t)=x(t)-\bar{X}(t)\rightarrow 0,~as~ t\rightarrow\infty$,
and $\bar{x}(t)$ (\underline{not $s(t)$}) is the synchronized state.

\begin{figure}
\centering\includegraphics[width=.5\textwidth]{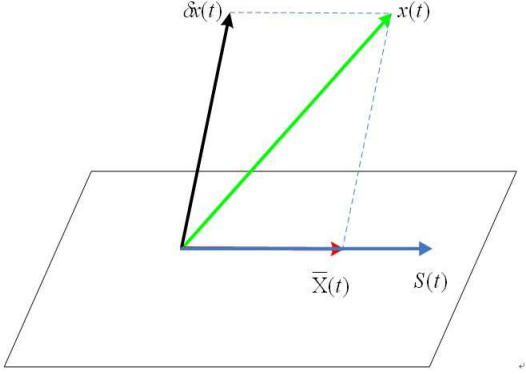}
\caption{Decomposition of $x(t)$} \label{div1}
\end{figure}

Let $\delta \bar{x}(t)$~be the variation near $\bar{x}(t)$, and $\delta
\bar{u}(t)=\Phi\delta \bar{x}(t)=[\delta
\bar{u}^{1}(t)^{\top},\cdots,\delta \bar{u}^{N}(t)^{\top}]^{\top}$, then we
have (see \cite{Lu2})
\begin{eqnarray}
\dot{\delta} \bar{u}(t)=[I_{N}\otimes DF(\bar{u}(t))]\delta
\bar{u}(t)+c\left(\Lambda\otimes\Gamma\right)\delta
\bar{u}(t)\label{chen1b}
\end{eqnarray}
and
\begin{eqnarray}
\dot{\delta}\bar{u}^{k}(t)=[D f(\bar{x}(t))+\lambda_{k}\Gamma]\delta
\bar{u}^{k}(t),~~k=1,\cdots,N
\end{eqnarray}

Different from
\begin{eqnarray}
\dot{\delta}{u}^{1}(t)=[D f(s(t))+\lambda_{k}\Gamma]\delta
{u}^{1}(t)\ne 0
\end{eqnarray}
here, due to $\delta\bar{u}^{1}(t)=0$, we have
\begin{eqnarray}
\dot{\delta}\bar{u}^{1}(t)=[D f(\bar{x}(t))+\lambda_{1}\Gamma]\delta
\bar{u}^{1}(t)=0
\end{eqnarray}

Thus, we can give
\begin{prop} \cite{Lu2} Consider the dynamical network (\ref{synn2w}). Let
\begin{align}
0=\lambda_{1}>\lambda_{2}\ge \lambda_{3}\ge\cdots\ge \lambda_{N}
\end{align}
be the eigenvalues of its coupling matrix $A$. If the following $N-1$-dimensional linear time-varying systems
\begin{align}
\dot{w}(t)=(Df(\bar{x}(t))+c\lambda_{k}\Gamma)w(t)~~~k=2, \cdots, N
\label{csaw2}
\end{align}
are locally exponentially stable, then
\begin{eqnarray*}
\|x(t)-\bar{x}(t)\|\leq M e^{-\epsilon t}
\end{eqnarray*}
which implies $\bar{x}(t)$ is the synchronized state.

\end{prop}

\begin{re}
It can be seen that the right side of the following equations
\begin{eqnarray}
\dot{x}^{i}(t)=f(x^{i}(t))+c\sum\limits_{j=1}^{N}a_{ij}\Gamma
x^{j}(t)\quad i=1,\cdots,N
\end{eqnarray}
contains two terms. The coupling term $c\sum\limits_{j=1}^{N}a_{ij}\Gamma x^{j}(t)$ controls $x(t)-\bar{X}(t)$. It is clear that
the coupling term $c\sum\limits_{j=1}^{N}a_{ij}\Gamma x^{j}(t)$ does not
contain any message of the synchronized state $s(t)$, except the initial
values $x^{i}(0)$ are near $s(0)$. Therefore, the coupling term does not
play any role to make a unstable $s(t)$ turn to be stable. Moreover, there
are infinite ${s}_{\alpha}(t)$ satisfying
$\dot{s}_{\alpha}(t)=f(s_{\alpha}(t))$ with $s_{\alpha}(0)$ being near
$s(0)$. Which one is the stable synchronized state defined in \cite{Wang}
for the coupled system (\ref{synn2w})?

\end{re}

\begin{re}
A basic prerequisite condition using variation near $s(t)$ is that all
$x^{i}(t)$,~$i=1,\cdots,N$, must be close to $s(t)$. However, as stated
above, under the condition (\ref{csaw}), one can not prove that
$x^{i}(t)-s(t)\rightarrow 0$, $i=1,\cdots,N$. Therefore, variational
analysis near $s(t)$ can not be applied. In particular, it can not be used for chaotic
oscillators.

\end{re}

\section{Numerical examples}

In this section, we will give several examples to illustrate our claims.

\begin{exa}
Consider the following coupled system:
\begin{eqnarray}\label{li1}
\left\{
\begin{array}{ll}
\dot{x}^{1}(t)&=tanh(x^{1}(t))+(-x^{1}(t)+x^{2}(t))\\
\dot{x}^{2}(t)&=tanh(x^{2}(t))+(x^{1}(t)-x^{2}(t))
\end{array}\right.
\end{eqnarray}
where the coupling matrix is
 $A=\left[\begin{array}{cc}
-1&1\\
1&-1
\end{array}
\right]$. Its eigenvalues are $\lambda_{1}=0$ and $\lambda_{2}=-2$.
~$f(s)=tanh(s)$, and $s=0$ is the unique equilibrium for
$\dot{s}(t)=f(s(t))$.

It is clear that
\begin{align}\label{l1}
\dot{w}(t)=[Df(0)+\lambda_{2}]w(t)=-w(t)
\end{align}
is stable, and
\begin{align}\label{l2}
\dot{w}(t)=Df(0)w(t)=w(t)
\end{align}
is unstable.

Numerical simulation (Figure \ref{Si1}) shows even initial values
~$x^{1}(0)=0.01$, $x^{2}(0)=0.02$ are chosen very close to $0$. However,
$x^{1}(t)\nrightarrow 0$ and $x^{2}(t)\nrightarrow 0,$ as
$t\rightarrow\infty$. Therefore, only the stability of the system
\begin{align}
\dot{w}(t)=[Df(0)+\lambda_{2}]w(t)
\end{align}
can not make the coupled system (\ref{li1}) synchronize to the equilibrium
point $"0"$ of the uncoupled system $\dot{s}(t)=tanh(s(t))$.

\begin{figure}
\centering\includegraphics[width=.4\textwidth]{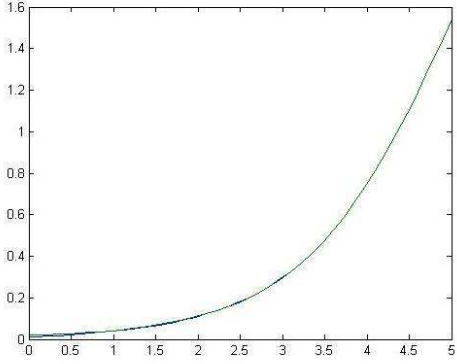} \caption{Synchronize
but does not converge} \label{Si1}
\end{figure}

On the other side, it is easy to see that $Df(\bar{x}(t))+\lambda_{2}<-1$.
Thus,
\begin{align}
\dot{w}(t)=(Df(\bar{x}(t))+\lambda_{2})w(t)
\end{align}
is stable. By Proposition 1 it can be concluded that
$x^{1}(t)-x^{2}(t)\rightarrow 0$.
\end{exa}

\begin{exa}
Consider following coupled system
\begin{align}
\left\{
\begin{array}{l}
\dot{x}^{1}(t)=f(x^{1}(t))+(-x^{1}(t)+x^{2}(t))\\
\dot{x}^{2}(t)=f(x^{2}(t))+(x^{1}(t)-x^{2}(t))
\end{array}\right.
\end{align}
where
\begin{align}
\left\{
\begin{array}{l}
f(x)=x-2r,~~x\in[2r-1,2r+1],~~r~is~\text{even} \\
f(x)=-(x-2r),~~x\in[2r-1,2r+1],~~r~is~\text{odd}
\end{array}\right.
\end{align}
and system $\dot{s}(t)=f(s(t))$ has multiple equilibria $\bar{s}=2r$.

It can be seen that
\begin{align}
\dot{w}(t)=[Df(0)+\lambda_{2}]w(t)=-w(t) \label{even1}
\end{align}
is stable, while
\begin{align}
\dot{w}(t)=Df(0)w(t)=w(t)\label{even2}
\end{align}
is unstable.

Simulation also shows that even $x^{1}(0)=0.05$,~$x^{2}(0)=0.15$ are very
close to $\bar{s}=0$, but when $t\rightarrow\infty$, $x^{1}(t)\nrightarrow
0$ and $x^{2}(t)\nrightarrow 0$. Instead,
\begin{align}
\dot{w}(t)=Df(2)w(t)=-w(t)\label{odd2}
\end{align}
is stable, $x^{1}(t)\rightarrow 2$ and $x^{2}(t)\rightarrow 2$. It means
that only the stability of the system
\begin{align}
\dot{w}(t)=[Df(0)+\lambda_{2}]w(t)=-w(t) \label{even1}
\end{align}
can not make the unstable equilibrium point $"0"$ of the uncoupled system
turn to be a synchronized state of the coupled system.

\begin{figure}
\includegraphics[width=.4\textwidth]{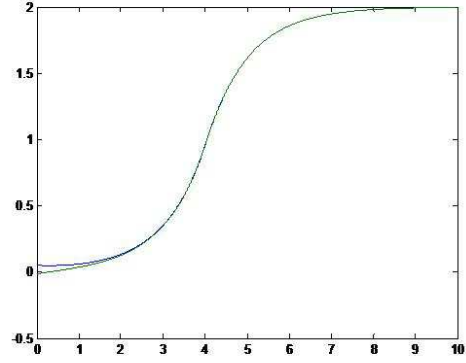}
\caption{Synchronize and does not converge to a unstable equilibrium but
converge to a stable equilibrium.}
\end{figure}

\end{exa}

The uncoupled system in the first example has a single equilibrium point
and in the second example has multiple equilibrium points. In case that the
equilibrium point $\bar{s}$ is not locally stable for the uncoupled system,
the trajectories $x^{i}(t)$, $i=1,\cdots,N$, of the coupled system
(\ref{synn2w}) will not converge to the equilibrium point (the synchronized
state $\bar{s}$ defined in \cite{Wang}).

In the following, we give a coupled system of chaotic oscillators to
illustrate our claims (see \cite{Lu2}). The initial values $x_{i}(0)$,
$i=1,\cdots,m,$ are assumed near $s(0)$. Simulations show that the coupled
system can reach synchronization, but the synchronized state is not the
trajectory of the uncoupled system $s(t)$.
\begin{exa}
Consider a coupled system with seven Chua's chaotic neural networks
\begin{eqnarray*}
\frac{dx^{i}}{dt}=-Dx^{i}(t)+T
g(x^{i}(t))+\sum\limits_{j=1}^{7}a_{ij}x^{j}(t),\quad i=1,\cdots,7
\end{eqnarray*}
here, $x^{i}=(x^{i}_{1},x^{i}_{2},x^{i}_{3})^{\top}\in R^{3}$, ~$D=I_{3}$,
$$T=\left[\begin{array}{ccc}
1.2500&-3.200&-3.200\\
-3.200&1.1000&-4.4000\\
-3.200&4.4000&1.000
\end{array}
\right]$$ $g(x^{i})=(g(x^{i}_1),g(x^{i}_2),g(x^{i}_3))$,
$g(s)=(|s+1|-|s-1|)/2$.~$A=(a_{ij})$, where
\begin{eqnarray*}
a_{ij}=\left\{\begin{array}{ll}1&i\ne j\\
-6&i=j\end{array}\right.\quad for~i=1,2,\cdots,7
\end{eqnarray*}
$s(t)$ is a solution of uncoupled system with initial value
$s(0)=[0.1,0.1,0.1]^{T}$.

The initial value for the coupled system are assumed to be
$x^{i}_{j}(0)=0.1+\delta{x}^{i}_{j}(0)$, where $\|\delta{x}^{i}(0)\|\le
0.01$,~$i=1,2,\cdots,7$.

Define $ K=\frac{1}{7}\sum\limits_{i=1}^{7}<\|x^{i}(t)-\bar{x}(t)\|> and
~W=\frac{1}{7}\sum\limits_{i=1}^{7}<\|x^{i}(t)-s(t)\|> $ where ~$<\cdot>$
denotes average with time.

\begin{figure}
\centering\includegraphics[width=.4\textwidth]{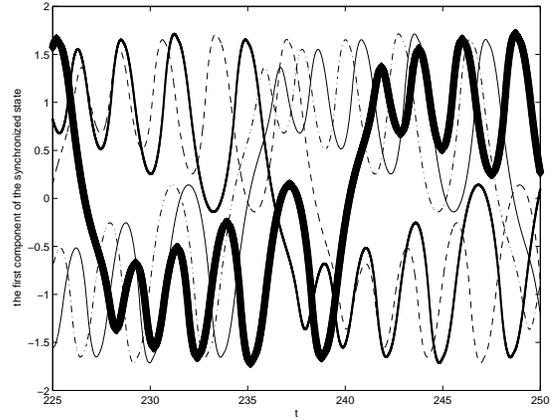}
\caption{Synchronized trajectories with different initial values
}
\label{chao1}
\end{figure}

\begin{figure}
\centering\includegraphics[width=.4\textwidth]{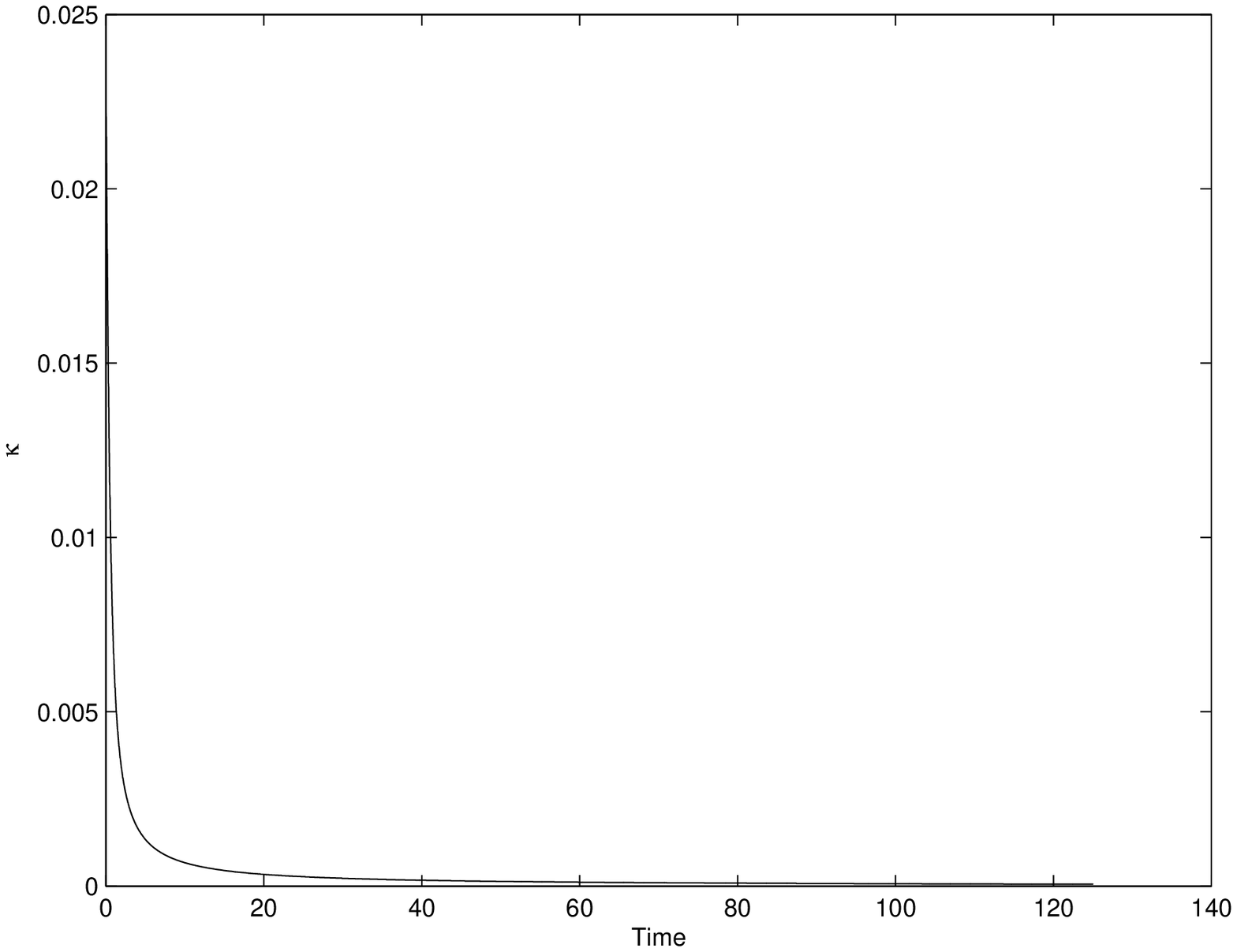}
\caption{Variation of $K$ with time}\label{chao2}
\end{figure}

\begin{figure}
\centering\includegraphics[width=.4\textwidth]{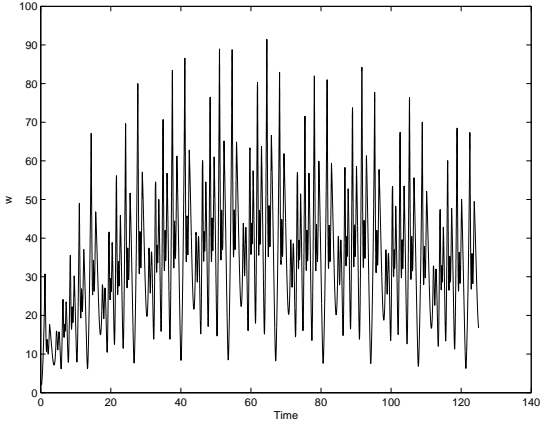}
\caption{Variation of $W$ with time}\label{chao3}
\end{figure}

Figure \ref{chao1} shows the first component of the different synchronized
states with different perturbations. It is clear that the synchronized
states heavily depend on the initial value, small perturbation of initial
value leads to serious change of the synchronized states. Figure
\ref{chao2}  shows that $K$ converges to 0, which means that the
synchronization manifold is stable; Instead, Figure \ref{chao3} shows that
$W$ does not converges to zero, which means that $x^{i}(t)-
s(t)\nrightarrow 0$. Therefore, even $x^{i}(0)$ are very close to $s(0)$
and the coupled system can synchronize, but $s(t)$ is not the synchronized
trajectory defined in \cite{Wang}.
\end{exa}

\section{Conclusions}

In summary, we conclude
\begin{itemize}
\item
The authors of \cite{Wang} misunderstand the synchronization by considering
synchronization of linear coupled system as asymptotically stable of some
solution of uncoupled system.

\item
It can be seen (see the Figure \ref{div1}) that
$$x(t)-S(t)=[x(t)-\bar{X}(t)]+[\bar{X}(t)-S(t)]$$
From previous derivation, the stability of following $N-1$-dimensional linear time-varying systems
\begin{align}
\frac{dw(t)}{dt}=(Df(\bar{x}(t))+c\lambda_{k}\Gamma)w(t)~~~k=2, \cdots, N
\label{csaw2}
\end{align}
leads to $x(t)-\bar{X}(t)\rightarrow 0.$ i.e., the coupled system (\ref{chen}) can reach synchronization and the synchronized state is $\bar{X}(t)$. That means that the coupling term in (\ref{synn2w} or \ref{chen}) (the eigenvalues $\lambda_{2},\cdots,\lambda_{N}$) is used to control $x(t)-\bar{X}(t)$. And the stability of the following system
\begin{align}
\frac{dw(t)}{dt}=Df(s(t))w(t)
\end{align}
leads to $\bar{X}(t)-S(t)\rightarrow 0$.

The condition that $N-1$ systems
\begin{align}
\dot{w}(t)=(Df(s(t))+c\lambda_{k}\Gamma)w(t)~~~k=2, \cdots, N
\end{align}
are stable can not lead to $x(t)-S(t)\rightarrow 0$.

\item
The synchronized state $\bar{X}(t)$ depends on initial value $x(0)$ heavily.
Any prescribed state $\dot{s}(t)=f(s(t))$ is never asymptotically stable
for the coupled system, unless $\dot{s}(t)=f(s(t))$ is asymptotically
stable itself.

\item
There are three possibilities of the dynamical behaviors for the uncoupled system $\dot{x}(t)=f(x(t))$:
\begin{enumerate}
\item
$\dot{s}(t)=f(s(t))$ is asymptotically stable, then under very mild
condition (for example, $\Gamma=I_{n}$), for the coupled system
(\ref{synn2w})
$$x^{i}(t)-s(t)\rightarrow 0,~~i=1,\cdots,N. $$

\item
$f=0$, and the system $\dot{s}(t)=0$ is neutral stable. For any initial
value $x^{i}(0)$, $i=1,\cdots,N$, $x^{i}(t)$ converge to a consensus
$\frac{1}{N}\sum_{i=1}^{N}x^{i}(0)$. But this consensus value is also
neutral stable. It is not asymptotically stable. Small perturbation of the
initial value will make the different consensus value and will never
return.

\item
$\dot{s}(t)=f(s(t))$ is unstable, in particular, it is chaotic, any
prescribed solution $s(t)$ of the uncoupled system $s(t)=f(s(t))$ is not a
synchronized state for the coupled system (\ref{synn2w}).

\end{enumerate}
\end{itemize}

\end{document}